# Emergence of low–temperature glassy dynamics in Ru substituted non–magnetic insulator $CaHfO_3$

Gurpreet Kaur and K. Mukherjee

School of Basic Sciences, Indian Institute of Technology Mandi, Mandi 175005, Himachal Pradesh, India

**Abstract**

Non–magnetic insulators/semiconductors with induced magnetism introduced via transition metal substitution are one of the promising materials in the field of spintronic, magnetoelectronics and magneto–optical devices. In this context, here, we focus on magnetism induced in a non–magnetic insulator $CaHfO_3$, by the substitution of 4d element Ru, at Hf–site. Structural investigations indicate that substitution of $Ru^{4+}$ (up to 50%) does not affect the original crystal structure of the parent compound. Magnetic studies divulge a crossover from a diamagnetic to paramagnetic state with 20% Ru substitution. Further replacement of Hf results in a glassy magnetic state in $CaHf_{1-x}Ru_xO_3$ ($0.3 \leq x \leq 0.5$). The nature of the low temperature glassiness (below 20 K) in these compositions is confirmed through Vogel–Fulcher and Power law, along with, magnetic memory effect and relaxation dynamics. The observed glassiness is explained through the phenomenological "hierarchical model". Our studies indicate that the presence of competing short range interactions among randomly arranged Ru cations in non–magnetic insulator $CaHfO_3$ are responsible for the observed low temperature magnetic state in this series with compositions > 0.25.

## 1. Introduction

Transition metal–based perovskites and their derivatives have been extensively studied over the last couple of decades due to the various novel properties like colossal magnetoresistance, large magnetocaloric effect, magnetization reversal, multiferroicity, etc. exhibited by them [1–5]. Dilute magnetic materials (belonging to the above family) in which ferromagnetism is induced by substitution of magnetic impurities, has recently gained substantial attention because of additional spin functionality [6–7]. Such materials are desirable candidates for spintronics, magneto–electronics, magneto–optics devices etc., due to the interplay of charge and spin degrees of freedom. Dilute magnetic alloys formed by II–VI and III–V elements are also investigated for these phenomena [8–10]. Generally, for all these cases, the magnetism has been induced by a low percentage of substitution. Further, the non–magnetic perovskites also provide a vast playground to explore and manipulate the physical properties of dilute magnetic oxides (DMO). Fe/Cr/Co doped $SrTiO_3$, Co doped $La_{0.5}Sr_{0.5}TiO_3$, Co/Mn doped $KTaO_3$ and $BaTiO_3$ etc. are examples of DMOs exhibiting room temperature ferromagnetism [11–13]. Similarly, enhancement in thermoelectric properties at high temperatures is predicted in Pr/Co/Nb doped $SrTiO_3$ [14,15]. $BaSnO_3$ is transparent to visible light and exhibits room temperature ferromagnetism. Ru doping at the Sn–site makes it desirable candidate for the magneto–optical devices [16]. The observed magnetism in these systems has been mostly explained by Ruderman–Kittel–Kasuya–Yosida (RKKY) interactions among magnetic cations via conduction electrons. Some of these materials, along with their insulating nature, also show glassy magnetic phases at low temperatures [17–18]. Initially, spin glass (SG)/ cluster glass (CG) dynamics had been anticipated in metallic alloys due to RKKY interactions [19]. But with time it was observed that several systems having semiconducting or insulating state show a glassy magnetic behaviour. This observed feature was explained through short range super–exchange interactions, and it was ascribed that these systems belong to a different universality class [20–21]. Additionally glassy magnetic state has also been induced in other non–magnetic perovskite insulators such as Sr, Nb and Mn substituted $LaCoO_3$ [22–24].

In this context, a non–magnetic insulator $CaHfO_3$ can be interesting. This compound has a wide bandgap of 6.4 eV and a high dielectric constant. Insulating to half metallic transition has been theoretically predicted in this compound due to the presence of higher concentration of O-defects [25]. Furthermore, it is optically transparent in the range of visible to deep ultraviolet light, making it a suitable candidate for optoelectronic devices [26–28]. Also due to its high melting point, it can be used in electrochemical devices. This compound also acts as a most promising host for scintillators due to the high atomic number of Hf [29–30]. However, the evolution of the physical properties of this compound due to substitution of magnetic impurity has not been explored. Another analogous perovskite compound, $CaRuO_3$, is metallic and isostructural to $CaHfO_3$. The magnetic state of $CaRuO_3$ is enigmatic because even after the presence of a magnetic atom, long–range ordering is absent. It happens due to the presence of lattice distortion [31–32]. The slow dynamic of spins at low temperature in this compound is confirmed by studying the magnetic state of the compounds via doping at Ru–site and by applying stress [33–39]. Since both these compounds have same structure, it is anticipated that $CaHf_{1-x}Ru_xO_3$ display full a solubility for whole x and presence of magnetic impurity will result in an evolution of novel magnetic properties in $CaHfO_3$.

In view of the above, in this manuscript, we have investigated the structural, magnetic, and thermodynamic properties of Ru doped $CaHf_{1-x}Ru_xO_3$ (x = 0 – 0.5) series through X–ray diffraction (XRD), dc and ac susceptibility, and heat capacity. Studies are limited to x = 0.5 composition as for x ≥ 0.60, superstructure of calcium hafnate starts developing, in contrast to our expectations. Physical properties of $CaRuO_3$ compound are also added here for comparison. The analysis of dc and ac susceptibility results indicate the absence of long–range magnetic ordering in these compounds, along with the observation of low temperature magnetic–glass state in x = 0.3, 0.4, and 0.5 compounds. Nature of the glass–dynamics investigated via different means indicates CG behaviour in these compounds. The magnetic properties below magnetic glass transition temperature have also been studied via time evolution of isothermal remanent magnetisation and memory effect as well. Our studies show the evolution of magnetic CG state due to competing short range magnetic interactions among randomly distributed Ru cations in the respective compositions of the insulating $CaHf_{1-x}Ru_xO_3$ series.

## 2. Experimental details

The polycrystalline compounds $CaHfO_3$ (Ru_0.0), $CaHf_{0.8}Ru_{0.2}O_3$ (Ru_0.2), $CaHf_{0.75}Ru_{0.25}O_3$ (Ru_0.25), $CaHf_{0.7}Ru_{0.3}O_3$ (Ru_0.30), $CaHf_{0.6}Ru_{0.4}O_3$ (Ru_0.4), $CaHf_{0.5}Ru_{0.5}O_3$ (Ru_0.5) and $CaRuO_3$ (Ru_1.0) are prepared using standard solid–state reaction method. The stoichiometric quantities of materials (purity > 99.9%) $CaCO_3$, $HfO_2$, and $RuO_2$ are mixed and grinded, followed by calcination for 12 hours at 1000° C. The calcinated mixture is regrinded to fine powder, pressed into pellets, and sintered at 1400° C for 48 – 60 hours. To investigate the crystal structure, powder XRD pattern was collected at room temperature in the range 10 – 90° with 0.02° step size, using Rigaku Smart Lab diffractometer with Cu–Kα source (λ = 1.54 Å). Rietveld refinement of the obtained XRD patterns is performed using Fullprof Suite software. The room temperature X–ray Photoelectron spectroscopy (XPS) was performed using NEXSA surface analysis model by Thermo Fisher scientific. The dc and ac magnetization measurements are performed on a Magnetic Properties Measurement System (MPMS) from Quantum Design Inc., USA. For heat capacity measurement, Physical Properties Measurement System (PPMS) from Quantum Design Inc., USA was used.

## 3. Results and discussions

### 3.1 Structural Analysis

#### 3.1.1 X–Ray Diffraction

Fig. 1 shows the Rietveld refined XRD pattern of $CaHf_{1-x}Ru_xO_3$ series. The Rietveld refinement has been done by using Fullprof software. For this linear interpolation between the set point of the background, and pseudo-voigt function for peak fitting have been used. All compounds crystallize in a single–phase orthorhombic structure with space group *Pnma*. In Ru_0.4 and Ru_0.5 compounds, high–intensity peaks of XRD patterns are accompanied by smaller peaks as can be seen in Fig. 1 (h). In this figure the peak at 2θ ~ 23° of Ru_0.5 compound is accompanied with the minor (extra) peak at higher side of 2θ ~ 24°. This minor peak is close to the peak observed in Ru_1.0 (~ 23.12°). Similar observations are noted for other peaks of Ru_0.5 and Ru_0.4 compounds. Thus, we have initially used the lattice parameters of Ru_1.0 for second phase which has same structure to Ru_0.0 to fit these peaks. Except the lattice parameters, all the other parameters remained same in both phases. Lattice parameters, volume of the unit cell, percentage of both phases, along with goodness of fit parameters are listed in Table 1. The percentage of developed second phase is less than 5% (Table 1). The lattice parameters of the second phase for Ru_0.4 and Ru_0.5 compounds are closer to the Ru_1.0 crystal structure. On exceeding Ru concentration above 50%, peaks of the superstructure $Ca_6Hf_{19}O_{44}$ have been observed. Hence, we have restricted our study to compositions up to $CaHf_{0.5}Ru_{0.5}O_3$ (i.e., Ru_0.5). Fig. 1 (h) also shows the gradual shifting of peaks of doped compounds from Ru_0.0 to Ru_1.0. This observation implies that Ru ions (with ionic radius 0.62 Å) replace the Hf ions (with ionic radius 0.71 Å). This observation is also in accordance with the decrement in lattice parameters and unit cell volume (shown in inset of Fig. 1 (g)) as we move from Ru_0.0 to Ru_1.0.

### 3.1.2 X–Ray Photoelectron Spectroscopy

For a proper analysis of the magnetic state of a compound, one should have an idea about the valence states of the constituent ions. Therefore, we have performed the photoelectron spectroscopy on selected compounds i.e., on the end members and one of the substituted compounds. As we wanted to check the valence state of Ru and Hf in each other's surroundings, we have chosen Ru_0.5 compound. Fig. 2 represents the core level XPS spectra of O 1$s$, Hf 4$f$, and Ru 3$p$, for these compounds. The inelastic background of the spectrum was subtracted using the Touggard method and then the peaks were fitted with the Voigt function. All compounds show two major peaks around 529.98 and 531.53 eV in the photoelectron spectrum of O 1$s$ (Fig. 2(a)–(c)). The 529.98 eV peak corresponds to $O^{2-}$ lattice oxygen while the latter resembles the surface chemisorbed oxygen or the oxygen–vacancy. On fitting XPS spectra of O 1$s$, one extra weak peak around 532.8 eV is obtained which corresponds to physically adsorbed oxygen species on the surface of compounds. A similar peak was also reported in another perovskite compound $LaFeO_3$ [40–41]. Fig. 2 (d) and (e) show the XPS spectrum of Hf cation in Ru_0.0 and Ru_0.5 compounds respectively. The Hf spectrum of Ru_0.0 shows two noticeable peaks at 16.77 eV and 18.44 eV, which are allocated to spin–orbit splitting components Hf 4$f_{7/2}$ and 4$f_{5/2}$ respectively. The spin–orbit splitting was calculated to 1.6 eV, which confirms the 4+ valence state of the Hf cation. Another peak observed around 25 eV corresponds to the loss part. The Hf 4$f_{7/2}$ and 4$f_{5/2}$ peaks have been shifted to 17.61 eV and 19.37 eV respectively, in the presence of Ru surrounding in Ru_0.5 compound. Also, the asymmetry observed around 16.29 eV might arise from the O-vacancies present in the system. Ru 3$p$ spectrum for Ru_0.5 and Ru_1.0 compounds has been shown in Fig. 2 (f) and (g) respectively. Two peaks at 462.12 eV and 484.26 eV corresponding to 3$p_{3/2}$ and 3$p_{1/2}$ spin–orbit splitting states are noted in the Ru 3$p$ spectrum of Ru_1.0 compound, and these peaks are shifted to 462.84 eV and 484.90 eV respectively in Ru_0.5 compound. These peaks are accompanied by two satellite peaks, seen as humps on higher binding energy sides. It confirms the predominate existence of 4+ state of Ru cation [41]. Furthermore, the shifting of core–level spectrum of Hf 4$f$ and Ru 3$p$ to higher binding energy in the presence of each other's surroundings in the Ru_0.5 compound has been associated to the disordered arrangement of these elements in B site of perovskite, $CaHfO_3$.

**3.2 DC magnetization study**

The temperature dependent dc susceptibility ($\chi(T)$) for all the compounds of the series, measured under the zero–field (ZFC) and field–cooled (FC) protocols in presence of magnetic field of 100 Oe is shown in the Fig. 3 (a) – (f). Ru_0.0 shows the crossover from diamagnetic to paramagnetic behaviour around 100 K on decreasing the temperature (Fig. 3 (a)). Here, it is to be noted that the minor thermomagnetic irreversibility observed between the ZFC and FC curves is an artefact of the measurement. Like $HfO_2$ [42], this magnetic behaviour of Ru_0.0 is not an intrinsic property of the compound but arises from presence of oxygen vacancies, confirmed via XPS analysis (as mentioned in the previous section). Thus, local magnetic moment induced by oxygen defects is responsible for the low temperature paramagnetic state of non-magnetic Ru_0.0 compound. However, when 20% of Ru is substituted, the $\chi(T)$ curve shows a positive value in the temperature range 2 – 300 K (Fig. 3 (b)). The magnetisation grows as the temperature is reduced to 2 K, with no bifurcation between the ZFC and FC curves. This implies the absence of any magnetic ordering/spin freezing in this compound. Also, the observed magnetic susceptibility of Ru_0.2 is couple of orders larger than Ru_0.0 compound. Thus, contribution of oxygen defects can be neglected in this case. The presence of Ru cations in Ru_0.2 is sufficient to produce a paramagnetic state and suppresses the diamagnetic behaviour observed in the parent compound. As noted from Fig. 3 (d), on increasing the Ru concentration to 30%, a maxima at ~ 17 K ($T_f$) is observed in the ZFC curve, while FC curve continues to increase as the temperature is decreased. The bifurcation between ZFC and FC curves starts around the irreversible temperature ($T_{irr}$) ~ 53 K ($\gg T_f$). A similar kind of behaviour in $\chi(T)$ curves is observed for compounds with Ru_0.4 and Ru_0.5 (Fig. 3 (e) and (f)). Notably, the values of $T_f$ and $T_{irr}$, as well as the difference between them lessen as Ru concentration rises. These observed features indicate the presence of magnetic ordering, yet no feature or anomaly is noticed in heat capacity data (as discussed in supplementary data). This suggests that these compositions lack any kind of long–range ordering. This type of behaviour is not unusual, and it has also been observed in other insulating as well as semiconducting materials having magnetic glass state or in superparamagnetic system [22–24, 43–45]. As mentioned earlier, a paramagnetic state is present in Ru_0.2, while Ru_0.3 exhibits a magnetic anomaly at low temperature. Hence, to find the percolation threshold at which short range magnetic interactions starts dominating, we have prepared an intermediate compound Ru_0.25. The ZFC and FC curves of this compound as a function of temperature are shown in Fig. 3 (c). A clear bifurcation between ZFC and FC curves is noted at $T_{irr}$ ~ 66 K while a weak anomaly in the ZFC curve is noted around ~ 15 K. Such features might arise due to the local ordering of magnetic moments at higher temperatures as compared to the macroscopic ordering temperature. Additionally, the bifurcation between the ZFC and FC curves in the other end member, Ru_1.0 (Fig. S2 (a)), occurs at 85 K, which is consistent with previous studies [46].

Inset of Fig. 3 (b) represents modified Curie Weiss law fitting of $\chi^{-1}(T)$ using equation $\chi = \chi_0 + C/(T-\theta_p)$ (where symbols having their standard meanings), in the paramagnetic region for all compounds (except diamagnet Ru_0.0). All the parameters obtained from the modified Curie–Weiss law are given in Table 2. The negative value of the Curie–Weiss temperature ($\theta_p$) indicates the dominance of antiferromagnetic interactions in Ru_0.2 to Ru_0.5 compounds, and in Ru_1.0 (shown in section S2 of supplementary data). In all compositions, a S–type feature is noted in the $\chi^{-1}$ vs T curve below paramagnetic range. It indicates the presence of short–range ferromagnetic interactions as also reported in $CaRu_{1-x}Ti_xO_3$ series [46]. The temperature independent susceptibility ($\chi_0$) is found to be in order of $10^{-5}$ emu/mol–Oe. Also, the experimental value of magnetic moment for Ru_1.0 (~ 2.96 $\mu_B$) is close to the moment of $Ru^{4+}$ ion in its low spin ($S = 1$) state. The parameters obtained (shown in section S2 of supplementary data) for Ru_1.0 matched well with the literature [46]. On incorporating Ru atom at the Hf site of a non–magnetic matrix, the slope of $\chi^{-1}(T)$ curves i.e., Curie–Weiss constant (C) varies linearly with the Ru concentration. However, $\theta_P$ does not vary linearly with x, implying the strength of magnetic interactions among Ru cations does not follow any pattern. This is not in accordance with the disordered Heisenberg antiferromagnet, where $\theta_P$ and C values scale linearly with concentration of magnetic impurity in the system [47]. Also, the high value of $\theta_P$ in comparison with $T_f$ indicates to the presence of frustration caused due to the presence of disorder at B–site of the perovskite, which might be responsible for observed magnetic glass state [48]. The experimental value of magnetic moment matches with calculated value of magnetic moment $\sqrt{(x*\mu_{Theor}(Ru^{4+}))^2}$, implying that the valency of Ru cation is matched well with the results of XPS analysis.

To further investigate the low–temperature magnetic state, we have measured dc $\chi$ ($T$) of Ru_0.3, Ru_0.4 and Ru_0.5 compounds at different magnetic fields as shown in the respective insets of Fig. 3 (d) – (f). Both the bifurcation temperature between ZFC – FC curves ($T_{irr}$) and peak temperature ($T_f$) of ZFC curve reduces on increasing the applied magnetic field. Such kind of features has been reported in unconventional SG systems [18, 49–50]. For a conventional SG, it is noted, $T_{irr} \leq T_f$, and below this temperature, FC curve becomes independent of temperature [51]. However, some studies report a temperature dependent FC curve with minima after $T_{irr}$ or $T_f$, in conventional SGs [52, 53]. The observation of $T_{irr} \geq T_f$ has also been noted in insulating SG systems [44, 49]. Hence, to determine the nature of the low temperature complex magnetic state in our system, isothermal magnetisation and ac susceptibility measurement are carried out.

Isothermal magnetization as a function of magnetic field ($M$ ($H$)) at 2 K and 300 K for all the prepared compounds are represented in the Fig. 4 (a) and (b). Within ±10 kOe, the $M$ ($H$) curve for Ru_0.0 (right panel of Fig. 4 (a)) show a linear dependence, but beyond this region, it shows the diamagnetic behaviour. The paramagnetic type of behaviour in Ru_0.0 compound at low temperature and low magnetic field observed in $\chi$ ($T$) and $M$ ($H$) curves might arise due to the presence of oxygen vacancies. In Ru_1.0 compound, the linear behaviour of $M$ ($H$) curve and negative value of $\theta_P$ indicates strong antiferromagnetic interactions (Fig. S2 of supplementary data). While $M$ ($H$) curves for Ru_0.2 and Ru_0.25 compounds at 2 K form S–like shape with no hysteresis at low field, or saturation at high field. On further diluting the Hf–site (0.30 ≤ x ≤ 0.5), we have noticed the opening of magnetic hysteresis in the low field region. However, magnetic saturation is not achieved even at 70 kOe. Such kind of $M$ ($H$) hysteresis without saturation even at 2 K is observed in systems having competing antiferromagnetic and ferromagnetic interactions. These competing magnetic interactions are responsible for glassy magnetic state or superparamagnetic behaviour or canted antiferromagnetism [22, 49–52]. The variation of coercive field and remanent magnetisation at 2 K with Ru concentration has been displayed in Fig. 4 (c). Increment in both parameters with Ru concentration (except Ru_1.0 due to its paramagnetic nature) have been observed, implying magnetic interactions in the system solely depend on the magnetic entity $Ru^{4+}$ cations. To get a better idea about the magnetic state, temperature response of the coercivity is noted. For this purpose, we have measured isothermal $M$ ($H$) curves at different temperatures below $T_{irr}$ for Ru_0.3, Ru_0.4 and Ru_0.5 compositions (Fig. 4 (d) – (f)). The substantial magnetic hysteresis has been noticed below the $T_f$ which reduces with temperature. The coercive field ($H_c$) for single domain particle i.e., superparamagnet varies as $T^{1/2}$, while it follows parabolic curve for canted antiferromagnets [5, 51, 54]. In magnetic glasses, the value of $H_c$ drops exponentially with temperature below $T_{irr}$, which is fitted with the following function [55]:

$$H_c = H_c(0) exp(-\alpha T) \tag{1}$$

where α is the temperature exponent. The respective insets of Fig. 4 (d) – (f) show the variation of $H_c$ with $T$ and fit using equation (1). The obtained parameters for all the three compounds are logged in Table 3. The exponential decrement of $H_c$ on increasing temperature has been reported for systems having slow dynamics of spins or group of spins [55–57]. This implies that Ru_0.3, Ru_0.4 and Ru_0.5 compounds have complex spin dynamics at low temperature.

### 3.3 AC susceptibility

AC susceptibility is a powerful tool to identify the glassiness in the system and its nature because the relaxation of spins/cluster of spins slows down in the glassy magnetic state on cooling the system from a paramagnetic region. The maximum relaxation $\tau = 1/f$ ($f$ is the frequency of applied $H_{ac}$) is obtained at freezing temperature $T_f$, thus, its value shifts to higher temperature on increasing the frequency. Therefore, we have measured temperature dependent ac susceptibility of compounds showing magnetic relaxation behaviour in the above context. Compounds Ru_0.2, Ru_0.25 and Ru_1.0 show no anomaly in ac susceptibility data and the value of ac susceptibility decreases with increasing temperature irrespective of frequency of applied ac magnetic field $H_{ac}$, indicating the absence of any type of magnetic ordering in these compounds (shown in Fig. S3 of supplementary data). Fig. 5 (a) – (c) presents the temperature dependent in–phase ($\chi'(T)$) and out–phase ($\chi''(T)$) components (in their respective insets) at different frequencies (13 – 931 Hz) of the ac magnetic field ($H_{ac}$) with amplitude of 1 Oe for Ru_0.3, Ru_0.4, and Ru_0.5 compounds respectively. A clear frequency dependent broad peak around $T_f$ is detected in the ac susceptibility (both in–phase and out–phase part) in terms of both peak position and peak intensity, indicative of a glassiness/slow spin dynamics in the system. Also, peaks with different frequencies tend to converge below $T_f$. A SG state appears when disorder or mixed exchange interactions give rise to an atomistic glassy phase with frozen spins below a well-defined freezing temperature. In contrary, in cluster glass (CG), the clusters having regularly arranged spins are frozen below the freezing temperature. Thus, this observed glassiness might arise from the interacting magnetic entities which are larger than atomic spins that constitute conventional SG system, as implied by the convergence of ac susceptibility curves at different frequencies below $T_f$ [58-60].

The ac susceptibility has also been measured at different applied dc fields ($H_{dc}$) with fixed $H_{ac}$ (1 Oe) of a single frequency (331 Hz) for Ru_0.3 compound, and Ru_0.4 and Ru_0.5 compounds Fig. 5 (d) – (f) shows the in–phase $\chi'$ ($T$) component of these compounds. Here the strong dependence of peak intensity on superimposed $H_{dc}$ is observed. On applying 100 Oe, the peak shifts towards the lower temperature side with a decrease in intensity. On further increasing $H_{dc}$, the peak changes to a broad hump and gets flattened. A similar feature has been observed in dc $\chi$ ($T$) on increasing the magnetic field. The occurrence of frequency and $H_{dc}$ dependent peak in ac susceptibility confirms the magnetic glass state in the system [54].

The variation of ac susceptibility with $H_{dc}$ could also be analysed by non–mean field theory (shown in insets of Fig. 5 (d) – (f)). So, we have fitted the field variation of inflection temperature calculated from d$\chi'$ ($T$)/d$T$ (not shown here) with the following equation [61]

$$T_f(H) = T_f(0)(1 - A * H^{2/\Phi}) \qquad (2)$$

where $A$ is the anisotropic strength parameter, $T_f$ (0) is the value of $T_f$ in absence of magnetic field and $\Phi$ represents the crossover exponent. This $\Phi$ exponent distinguishes between weak anisotropy regime and strong anisotropy regime. In strong anisotropic (strong irreversibility) regime, the $T$–$H$ phase transition line follows the Almeida–Thouless (AT) line with a value of $\Phi$ ~ 3. The AT line distinguishes the non-ergodic (i.e. SG) phase from the ergodic (i.e. paramagnetic) phase. While in a weak anisotropic regime, Gabay–Toulouse (GT) line is followed in the $T$–$H$ phase diagram, with a value of $\Phi$ ~ 1. For Ru_0.3 compound, obtained values of $\Phi$ ~ 2.059 and $T_f$ (0) = 19.0 ± 0.2 K. In the case of Ru_0.4 and Ru_0.5 compounds, values of $\Phi$ are increased to 4.55 and 4.097, and $T_f$ (0) = 20.0 ± 0.1 K and 17.2 ± 0.2 K respectively. The obtained values of $T_f$ (0) are very close to freezing temperature, as observed in frequency dependent data of $\chi'$ ($T$). From values of $\Phi$, it can be concluded that observed glassy behaviour in Ru substituted compounds follow the non–mean field model and belong to different universality class. Such kind of behaviour is also seen in other magnetic glasses like $Er_5Pd_2$, $Fe_{2-}$ $_x$Mn$_x$CrAl, Cr$_{0.5}$Fe$_{0.5}$Ga, etc [62–64]. To conclude, it can be said that a transformation from weak anisotropy to strong anisotropy is observed in this series on increasing the Ru concentration.

To further explore the spin/clusters dynamics, Mydosh parameter $\delta T_f = \Delta T/T_f\Delta\ln f$, indicating the shift of peak temperature in response of applying $H_{ac}$ of varying frequencies, is determined from $T_f$. The shifting of peak ($\delta T_f$) depends upon the interactions among the magnetic entities (spins/magnetic clusters). Thus, stronger the interaction among magnetic particles, weaker is the sensitivity of $T_f$ to the frequency of ac field. The values of $\delta T_f$ for canonical SG's like AuMn and CuMn is ~ 0.005, while for superparamagnets $\delta T_f$ ~ 0.5, due to very weak interactions among particles [51]. The values $\delta T_f$ for our compounds are given in Table 4. It is noted that the values lie in–between that of SG and superparamagnets. However, the $\delta T_f$ value is comparable to that reported CG's in metallic as well as in the insulating system (0.02–0.06) [58–66].

To further probe the spin dynamics of the system, the frequency dependent peak is analysed with the Vogel–Fulcher (V–F) law:

$$\tau = \tau_0 \exp\left(\frac{E_a}{k_B(T_f-T_0)}\right) \quad (3)$$

where $T_0$ is the V–F temperature (which tells about the strength of interaction between spins/clusters), $k_B$ is the Boltzmann constant, $E_a$ is the average thermal activation energy, and $\tau_0$ is the characteristic relaxation time. The scaling according to equation (3) is represented in Fig. 6 (a) – (c) for Ru_0.3, Ru_0.4 and Ru_0.5 compounds. The obtained values are presented in Table 4. The non–zero value of $T_0$ and agreement with V–F law indicates the presence of finite interactions, in contrast to superparamagnets or non–interacting spin dynamics [48, 51]. The presence of finite interactions has also been confirmed from comparable values of $E_a/k_B$ and $T_0$. It is to be noted that in weak interactions regime, the ratio of $E_a/k_B$ to $T_0$ is substantially lower than 1, whereas it is reverse for the strong interactions regime. Also, the obtained value of $\tau_0$ ~ $10^{-6}$ to $10^{-8}$ s, far greater than for conventional SG systems ($10^{-13}$ s), has also been reported in other insulating CG system [57]. The temperature dependent relaxation time is also analysed by Power law described as:

$$\tau = \tau^*\left(\frac{T_f}{T_g} - 1\right)^{-z\nu} \quad (4)$$

where $z$ is the dynamic exponent which describes the slowing down of relaxation, $\nu$ is the correlation length exponent, $\tau^*$ describes the flipping time for relaxing entities, and $T_g$ is the true SG temperature. Fig. 6 (d) – (f) shows the scaling of $\tau$ with reduced temperature $\varepsilon = (T_f/T_g-1)$. All the parameters are given in Table 4. For SG, $z\nu$ falls in the range of ~ 4 – 10, and $\tau^*$ falls in the range of $10^{-10} – 10^{-12}$ s. For CGs, it is reported that the value of $\tau^*$ lies in the range of $10^{-7} – 10^{-10}$ s [63]. In our case, even though the value of $z\nu$ lies in the range of SG's, the value of $\tau^*$ fall in the range reported for CGs. Thus, presence of Ru at B–site of non–magnetic insulator $CaHfO_3$ leads to formation of insulating CGs at low temperature. The origin of magnetic glassiness in insulators has been explained via the presence of competing magnetic super–exchange interactions [27] and the geometrical frustration in the system [55, 56]. Hence, it can be said that in Ru_0.3, Ru_0.4 and Ru_0.5 compounds the presence of competing magnetic interactions might be responsible for the CG formation. Similar kind of behaviour has also been reported in other disordered perovskites [23–24, 48, 57].

To understand the nature of magnetism in this series, the $T$ – x phase diagram, based on the results of dc and ac susceptibility have been plotted in Fig. 7. As mentioned before, the nature of magnetism in Ru_0.0 is diamagnetic. For x ≥ 0.25, paramagnetic phase noted above 100 K, whereas, Ru_0.2 is paramagnet to the lowest measured temperature of 2 K. The paramagnetic phase is presented by a grey region in the phase diagram (Phase I). On further decreasing the temperature below 100 K, the irreversibility between ZFC and FC curves noted, indicating toward the local ordering of magnetic moments in contrast to long ranged ordering. The irreversibility decreases on moving from Ru_0.25 to Ru_0.5. This phase between paramagnetic state and frozen state is represented by violet region (Phase II) in the phase diagram. The CG state has been observed at low temperatures for Ru_0.3 – 0.5 compounds (shown as green coloured region (Phase III)). The magnetic interactions among randomly distributed Ru cations in non-magnetic insulator Ru_0.0 are mainly antiferromagnetic, as implied from the negative value of $\theta_p$. In an insulator, generally the magnetic interactions are of super-exchange in nature. In Ru_0.25 compound, these interactions arise but are not strong enough to give rise to an ordered or frozen state. For this composition, the interconnected network of nearest neighbouring magnetic atoms at the atomic sites is not formed. However, this network becomes strong at higher concentration and a magnetic glassy state is noted. The observed glassiness could be explained through the random distribution of Ru cations and competing super-exchange interactions between nearest and next nearest neighbouring magnetic cations $Ru^{4+}$. The origin magnetic glass state from competing super-exchange interactions is not unusual. Such behaviour has also been observed in $Ca(Fe_{1/2}Nb_{1/2})O_3$ [48].

### 3.4 Aging effect and isothermal remanent magnetization

Since aging effect and isothermal remanent magnetisation are the features of magnetic glass system, we have carried out these measurements on Ru_0.3, Ru_0.4 and Ru_0.5 compounds and for comparison, also on Ru_0.25 compound. For the aging effect phenomenon, following protocol is carried out. The sample is cooled in ZFC mode from paramagnetic region to measuring temperature (2 K) and then the system is allowed to age for different waiting time ($t_w$) of 10 , 100 and 10,000 s. After that, a magnetic field of 500 Oe is applied and the time response of ZFC magnetization is noted (Fig. 8 (a) – (d)). The value of magnetization and its growth (except for x = 0.25), gets reduced on increasing $t_w$. This indicates the influence of aging on the size of magnetic entities and frozen energy barriers associated with the non-ergodic phase. This is an obvious feature for a glassy magnetic system, where spins/magnetic entities fall into a more stable and deeper energy valley on increasing the waiting time $t_w$ [65–66]. This age dependent phenomenon also confirms the presence of metastable states below $T_f$ in 30–50% Ru substituted compounds. On the other hand, Ru_0.25 compound also shows the aging effect but the difference between $M_{ZFC}(t)$ curves for $t_w$ = 10 s and 1000 s is very small. For the other members of this series which do not show any frequency dependent feature in ac susceptibility, this measurement was not carried out.

To further explore the low–temperature magnetic state of these compounds, isothermal remanent magnetic relaxation ($M_{IRM}(t)$) was performed at temperatures primarily below $T_f$. In this protocol, the system is again cooled in ZFC mode from paramagnetic state to the desired temperature, then a magnetic field greater than coercive field ($H_c$) is applied for 20 minutes. After switching off the field, the $M_{IRM}(t)$ is recorded as a function of time for 1800 s. Such protocol for $M_{IRM}(t)$ is recorded at different measuring temperatures 5, 10, and 20 K. In Fig. 8 (e) – (h), $M_{IRM}(t)/M_{IRM}(0)$ curves at different temperatures are shown for Ru_0.25 to Ru_0.5 compounds. Pattern followed by $M_{IRM}(t)/M_{IRM}(0)$ curves for Ru_0.25 compound is different from the other compounds, implying the absence of a significant glassy nature at low temperature. In Ru_0.3, Ru_0.4 and Ru_0.5 compounds, the slowing down of decay rate of the remanent magnetization on decreasing temperature is clearly observed. The magnetic relaxations at these temperatures are well fitted with the stretched exponential equation:

$$M_t(H) = M_0(H) + [M_\infty - M_0(H)][1 - \exp\left(-\frac{t}{\tau}\right)^\beta]  \quad (5)$$

Here $M_0(H)$ and $M_\infty(H)$ are magnetization values at $t \to 0$ and $t \to \infty$ respectively whereas $\tau$ is the characteristic relaxation time and $\beta$ is the stretching exponent. The value of $\beta$ depends upon the energy barriers involved for relaxation process. $\beta = 0$ indicate the absence of any relaxation in the system while $\beta = 1$ indicate relaxation with single time constant. The intermediate value of $\beta$ talks about the distribution of relaxation times due to the presence of multiple degenerate levels in the frozen state. All the parameters obtained from equation (5) are presented in Table 5. The value of $M_0(H)/M_\infty(H)$ approaches 1 on decreasing temperature. Also, the relaxation time $\tau$ increases on lowering the temperature (shown in inset of Fig. 8 (h)). The variation of $M_0(H)/M_\infty(H)$ and $\tau$ with temperature supports the slowing down of magnetic relaxation on decreasing temperatures and the presence of metastability below $T_f$. The value $\beta < 1$ indicates the presence of magnetic anisotropy in the system. Inset of Fig. 8 (h) shows very weak variation of $\beta$ with temperature. The value of $\beta$ is independent of Ru concentration. Also, the obtained values of $\beta \sim 0.5$, is close to the values reported in other CG compounds like $Zn_3V_3O_8$, $Nd_5Ge_3$, $Cr_{0.5}Ga_{0.5}Ga$, and insulating magnetic glass compounds $SrTi_{0.5}Mn_{0.5}O_3$, $Sr_{2-x}La_xCoNbO_6$ [18, 64-66]. The non–zero value of $\beta$ at 20 K (above $T_f$) suggests the presence of short–range interactions among magnetic entities even above the $T_f$. Thus, the results of aging effect and IRM magnetization also support the presence CG dynamics in Ru_0.3, Ru_0.4 and Ru_0.5 compounds.

### 3.5 Memory effect

The memory effect is also a characteristic feature of glassy system due to degenerate ground state and it provides insight about the spin dynamics of such systems. As a result, we investigated the memory effect and rejuvenation effect on Ru_0.3, Ru_0.4 and Ru_0.5 compounds using ZFC and FC protocols. The results of this measurement are shown in Fig. 9. For FC memory effect (Fig. 9 (a) – (c)), the protocol reported in Ref [68] is followed. Temperature dependent magnetization has been recorded from 300 – 2 K in the presence of 100

Oe, with temporary halts at $T_H$ = 25, 15, 8, and 5 K. At each halt temperature $T_H$, the magnetic field is turned off and waited for 2 hours. After each stop at waiting temperature, the field of 100 Oe is applied again, and the magnetization measurement under field cooled cooling (FCC) protocol is resumed. The obtained $M(T)$ curve is presented as $M_{FCC}^{stop}$. At the halt temperatures, magnetization of above–mentioned compounds decays. It rises again when the magnetic field is resumed, resulting in a step–like behaviour in $M_{FCC}^{stop}$ curve. When the system reaches 2 K, the $M(T)$ is measured immediately under FC warming protocol of the compounds without any stops. The obtained $M(T)$ curve is presented as $M_{FCW}^{mem}$ which also shows the smeared step–like anomalies at the halted temperatures except at 25 K. This behaviour of $M_{FCW}^{mem}$ indicates that the compounds (for Ru_0.3 to Ru_0.5) in their glassy state remember their earlier thermal history of magnetization. As $T_H$ = 25 K is much beyond the glass transition temperature $T_f$, no effect on $M_{FCW}^{mem}$ has been observed. The compounds are cooled down again in presence of 100 Oe magnetic field without any pauses and the FC warming $M(T)$ is recorded, presented as $M_{FCW}^{ref}$. The $M_{FCW}^{ref}$ curve does not show any anomaly at the previous halted temperatures, indicating the thermal history – memory effect of these compounds has an intrinsic origin. The FC memory effect has been reported for both superparamagnetic and magnetic glass systems [69–70]. Hence, to further scrutinize the nature of the low–temperature magnetic phase of the system, ZFC memory effect is performed on the above–mentioned compounds (shown in Fig. 9 (d) – (f)). The ZFC memory effect, which is a distinctive signature of SG's emerging from cooperative spin–spin interactions, it is absent in superparamgnetic systems [70–71]. In this procedure, the compounds are cooled to 2 K in absence of field with a 2–hours halt at temperatures of 25, 15 and 8 K. During aging, at the halt temperatures below $T_f$ (i.e. in non-ergodic state), the growth of magnetic entities and frozen energy barriers increases simultaneously. At temperature 2 K, magnetic field (100 Oe) is switched on and $M_{ZFC}^{stop}(T)$ was noted during warming the system to 300 K. For the reference curve, the system is again cooled down to 2 K in the absence of magnetic field without any halts and then $M_{ZFC}^{ref}$ curve is measured on warming the system in presence of 100 Oe field. At the halt temperature of 15 K, the deviation of $M_{ZFC}^{stop}$ curve from $M_{ZFC}^{ref}$ curve has been clearly observed due to aging at this temperature. This deviation vanishes on increasing the temperature in between the 15 K and $T_f$. Thus, the system rejuvenates above 15 K even in the non-ergodic phase. Insets of Fig. 9 (d) – (f) present the difference between these two curves $M_{ZFC}^{ref}$ - $M_{ZFC}^{stop}$ as a function of temperature of respected compounds with Ru_0.3,

Ru_0.4 and Ru_0.5. The curve $M^{\text{ref}}_{\text{ZFC}}$ - $M^{\text{stop}}_{\text{ZFC}}$ for all compounds clearly displays the dips at the halt temperatures of 8 and 15 K, implying the presence of interaction among the magnetic entities instead of independent relaxations of them giving rise to superparamagnetic behaviour. Thus, memory and rejuvenation effect are obtained under ZFC protocol. Again, there is no ZFC memory effect at the halted temperature of 25 K because it is above $T_f$. Such behaviour of memory effect under ZFC and FC procedures supports non–equilibrium characteristics of the magnetic glass dynamics at low temperature in the Ru substituted compounds from 30–50% concentration.

To further explore the memory effect, cooling and heating temperature cycling has been performed on magnetic relaxation of Ru_0.3, Ru_0.4 and Ru_0.5 compounds. The protocol described in Refs. [68, 71] was followed for this relaxation memory measurement.

For cooling $T$ cycle, the magnetic relaxation under ZFC and FC protocols are measured. Therefore, the system is cooled down to 8 K (below $T_f$) in absence of field, then the logarithmic increase of magnetization with time ($M(t)$) is recorded in the presence of 50 Oe for ZFC method for time $t_1 = 1800$ s, respectively. There after the compound is quenched to a lower temperature of 5 K in the presence and again $M(t)$ is recorded for second time $t_2 = 1800$ s. After that, the compound is heated back to 8 K temperature in the presence, then $M(t)$ is recorded for third time $t_3 = 1800$ s in presence of field. The upper panels of Fig. 10 (a) – (c) represents the ZFC relaxation during the cooling $T$ cycle for Ru_0.3, Ru_0.4 and Ru_0.5 compounds respectively. Now for FC mode, the system is field cooled (in 50 Oe) down to the temperature below $T_f$. On reaching the measuring temperature 8 K, magnetic field is switched off and then logarithmic decay of magnetisation as function of time is recorded for 1800 s at 8, 5 and 8 K respectively. The lower panels of Fig. 10 (a) – (c) shows the magnetic relaxation under FC mode of the cooling cycle. From the figures it is noted that: (i) relaxation of $t_3$ starts where relaxation of $t_1$ ends. This clearly indicates that these compounds recall their previous history i.e., before temporarily cooling down in both ZFC and FC cooling cycles and (ii) observation of weak relaxation at $t_2$ during immediate quenching to 5 K, implies the system does not get ample time to get relaxed at this temperature.

Similarly, to see the effect of memory during heating $T$ cycle, the above protocols for ZFC and FC mode are followed, and magnetic relaxation of the compounds has been noted for the sequence of temperatures 8, 11 and 8 K. The outcome of the measurements is represented in Fig. 10 (e) – (f) for Ru_0.3, Ru_0.4 and Ru_0.5 compounds (in FC mode and ZFC mode). It is observed that the heating $T$ cycle appears to erase the pervious memory effect and re–initialise the magnetic relaxation during temporary heating in both ZFC and FC situations. Thus, the system rejuvenates the magnetic relaxation during positive cycle.

The memory effect observed in magnetic relaxation of glassy state is generally described by either Droplet model [68, 72] or Hierarchical model [73–74]. In droplet model, the overlap length $l_{\Delta T}$ is introduced which describes the critical length scale at which correlation of spins at two different temperatures (below magnetic glass transition temperature) is same. Thus, one expects the symmetrical response of magnetic relaxation irrespective $T$ cycles. However asymmetric response of magnetic relaxation is expected for Hierarchical model. In Hierarchical model, the system has multi–valley energy landscape, which corresponds to metastable configurations that existed at temperatures below $T_f$, and these valleys split into sub valleys on decreasing temperature. The magnetic relaxation of Ru_0.3, Ru_0.4 and Ru_0.5 compounds during cooling and heating $T$ cycle follows the Hierarchical model i.e., when the temperature of these compounds is reduced from 8 K to 5 K, the system becomes locked in sub–valleys or metastable states. If the temperature changes significantly, the energy barriers between primary valleys become too high, and the system does not have enough time to overcome these barriers and get relaxed within the sub valley. As a result, during the cooling $T$ cycle, weak relaxation is obtained at a temporarily halted temperature of 5 K. On returning to 8 K these sub–valleys merge to form the original valleys and the system restores its previous memory. However, in the heating $T$ cycle, when the system returns to temperature 8 K from temporary halt at 11 K, the value of magnetic relaxation has been re–initialised. This is because valleys at 8 K merge to form super–valley or the energy barriers among these valleys reduced on increasing temperature to 11 K. When the temperature is dropped once again to 8 K, the super – valley at 11 K restored to the original valleys at 8 K but the relative occupancy of each valley has been changed. Thus, observation of memory effect in cooling temperature cycle and re–initialisation in heating temperature cycle in our compounds supports the hierarchical arrangement of metastable states in glassy phase. This observation in magnetic relaxation as well as analysis of dynamic susceptibility confirms the presence of magnetic interactions among the clusters of spins rather than independent behaviour of individual spins.

## 4  Conclusions

In conclusion, we have investigated the physical properties of a series of insulating perovskite $CaHf_{1-x}Ru_xO_3$ ($0 \leq x \leq 0.50$). Our studies reveal that long range magnetic interaction is absent across the series. For the $0.30 \leq x \leq 0.50$ compounds, magnetic glass dynamics at low temperature is confirmed through both static and dynamic susceptibility. Further investigations reveal the cluster glass nature of the glassy phase. The asymmetric response of magnetic relaxation during heating and cooling $T$ cycles implies that the energy landscape picture of this glassy phase is described in terms of hierarchical model. Our study indicates that the competing short–range interactions among the inhomogeneous distributed magnetic $Ru^{4+}$ cations in non–magnetic matrix $CaHfO_3$ is responsible for the low temperature cluster glass dynamics in this insulating series.

**Acknowledgements**

The authors acknowledge IIT Mandi for the experimental facilities and financial support.

**Data availability statement**

All data that support the findings of this study are included within the article.

Table 1 Structural parameters obtained from Rietveld refinement of XRD pattern using Fullprof suit Software

| Compounds | | a (Å) | b (Å) | c (Å) | V (Å$^3$) | Phase % | $\chi^2$ |
|---|---|---|---|---|---|---|---|
| Ru_0.0 | | 5.733 (1) | 7.980 (0) | 5.570 (1) | 254.83 (1) | 100 | 3.08 |
| Ru_0.2 | | 5.692 (1) | 7.935 (1) | 5.546 (2) | 250.48 (1) | 100 | 2.21 |
| Ru_0.25 | | 5.681 (0) | 7.923 (1) | 5.540 (0) | 249.34 (2) | 100 | 2.69 |
| Ru_0.3 | | 5.686 (1) | 7.921 (1) | 5.541(1) | 249.55 (4) | 100 | 2.58 |
| Ru_0.4 | Phase1 | 5.684 (1) | 7.918 (1) | 5.536 (1) | 249.14 (3) | 98.29 | 2.33 |
| | Phase2 | 5.542 (0) | 7.670 (4) | 5.408 (6) | 229.90 (4) | 1.71 | |
| Ru_0.5 | Phase1 | 5.672 (0) | 7.909 (1) | 5.536 (0) | 248.34 (3) | 95.24 | 1.95 |
| | Phase2 | 5.561 (1) | 7.717 (3) | 5.358 (4) | 230.00 (2) | 4.76 | |
| Ru_1.0 | | 5.528 (1) | 7.663 (1) | 5.359 (1) | 227.02 (1) | 100 | 1.97 |

Table 2 Parameters obtained from Curie–Weiss fit on $\chi^{-1}$ vs $T$ at an applied field of 100 Oe

| Compounds | $T_f$ (K) | $\chi_0 *10^{-5}$ (emu/mol/Oe) | $\theta_P$ (K) | $\mu_{eff}$ ($\mu_B$) | $\mu_{theor}$ ($\mu_B$) |
|---|---|---|---|---|---|
| Ru_0.2 | - | 3.32 (3) | -84.9(2) | 1.25 | 1.26 |
| Ru_0.25 | 23.58 | 6.45 (1) | -194.2(1) | 1.46 | 1.41 |
| Ru_0.3 | 17.00 | 3.53 (1) | -111.4(1) | 1.58 | 1.54 |
| Ru_0.4 | 17.80 | 3.29 (3) | -35.8(1) | 1.51 | 1.78 |
| Ru_0.5 | 14.57 | 3.52 (2) | -65.2(1) | 1.98 | 2.00 |

Table 3 Parameters obtained from equation (1) fitted on coercive field $H_c$ vs $T$

| Compounds | $H_0$ (kOe) | $\alpha$ (K$^{-1}$) |
|---|---|---|
| Ru_0.3 | 3.52 (22) | 0.21 (2) |
| Ru_0.4 | 8.96 (13) | 0.33 (1) |
| Ru_0.5 | 3.43 (22) | 0.23 (2) |

Table 4 Mydosh parameters and parameters obtained from Vogel–Fulcher and power law fitted on $T_f$ obtained from ac susceptibility

| Compounds | $T_f$ (K) | $\delta T_f$ | Vogel–Fulcher law | | | Power law | | |
|---|---|---|---|---|---|---|---|---|
| | | | $T_0$ (K) | $\frac{E_a}{k_B}$ (K) | $\tau_0$ (s) $*10^{-7}$ | $T_g$ (K) | $z\nu$ | $\tau^*$ (s) $*10^{-8}$ |
| Ru_0.3 | 17.89 | 0.036 | 16.0 | 21 (1) | 10.00 (0) | 16.9 | 5.6 (2) | 1.10 (1) |
| Ru_0.4 | 19.29 | 0.028 | 17.7 | 18 (1) | 12.50 (1) | 18.6 | 4.9 (3) | 1.00 (1) |
| Ru_0.5 | 16.50 | 0.030 | 14.0 | 39 (1) | 0.11 (2) | 15.5 | 6.4 (1) | 0.14 (4) |

Table 5 Parameters obtained from stretched exponential formula using equation (5) on $M_{IRM}$ vs $t$

| Compounds | $T$ (K) | $M_\infty/M_0$ | $\beta$ | $\tau$ (s) |
|---|---|---|---|---|
| | 5 | 0.931 | 0.465 (1) | 540 (2) |
| Ru_0.3 | 10 | 0.830 | 0.477 (1) | 493 (2) |
| | 20 | 0.810 | 0.480 (1) | 403 (2) |
| | 5 | 0.925 | 0.498 (1) | 529 (3) |
| Ru_0.4 | 10 | 0.833 | 0.490 (1) | 524 (3) |
| | 20 | 0.778 | 0.482 (1) | 420 (2) |
| | 5 | 0.927 | 0.484 (1) | 619 (6) |
| Ru_0.5 | 10 | 0.845 | 0.478 (1) | 487 (3) |
| | 20 | 0.807 | 0.476 (1) | 398 (2) |

**Figures:**

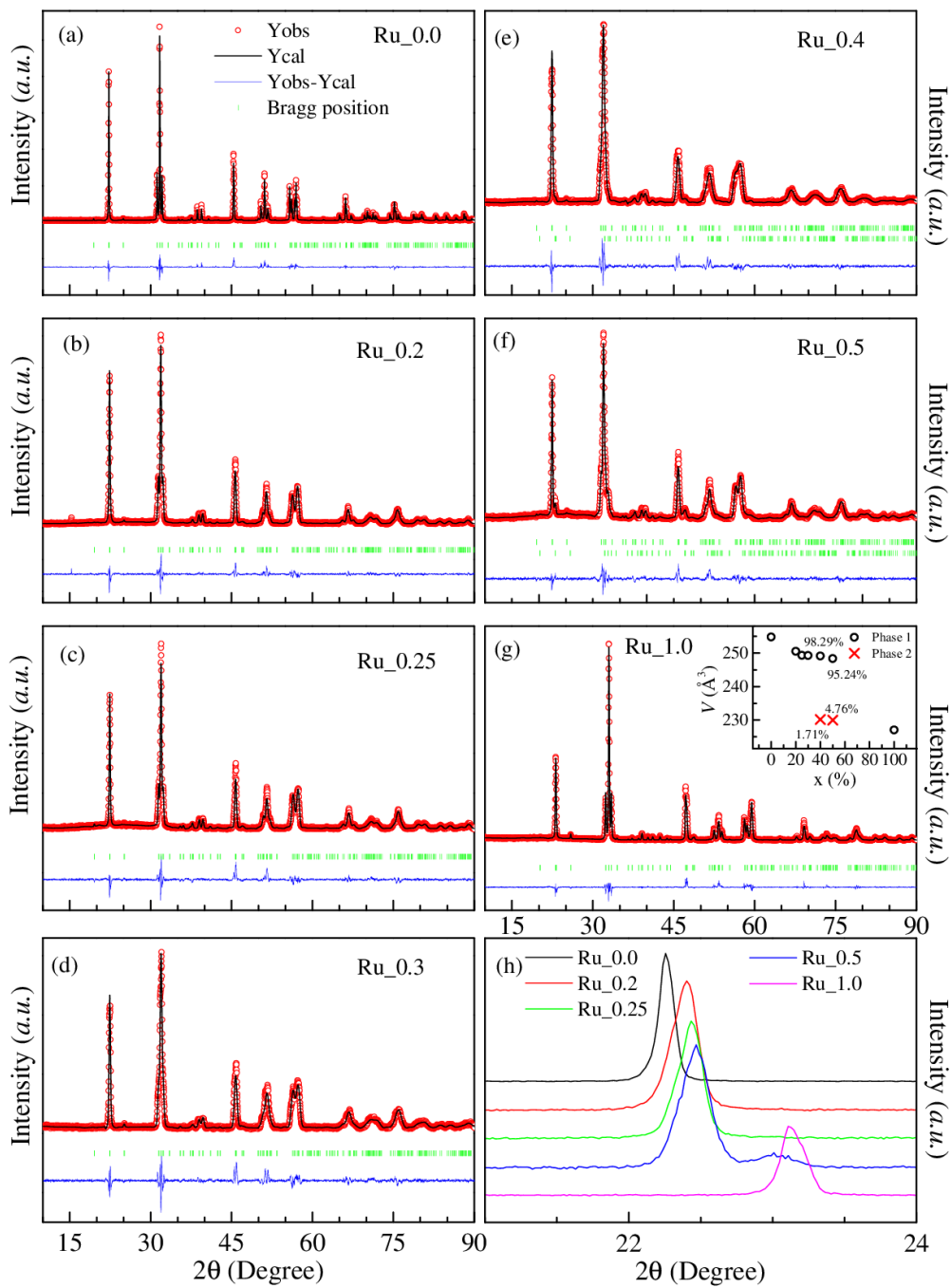

Figure 1 Rietveld refined X–ray diffraction pattern of CaHf$_{1-x}$Ru$_x$O$_3$: (**a**) Ru_0.0 (**b**) Ru_0.2 (**c**) Ru_0.25 (**d**) Ru_0.3 (**e**) Ru_0.4 (**f**) Ru_0.5 (**g**) Ru_1.0 and (**h**) shows the shifting of peaks with increasing Ru doping. Inset of (**g**) shows the variation of unit cell volume of these compounds.

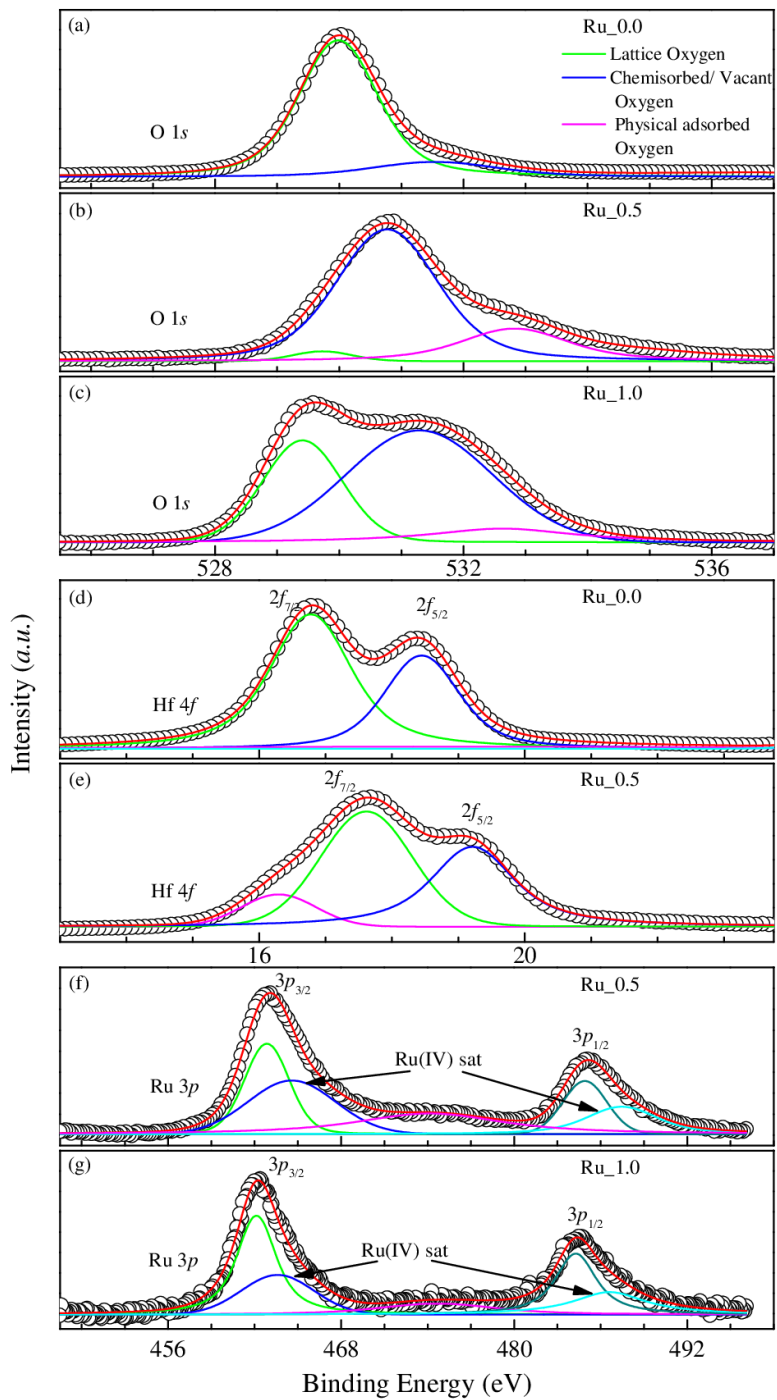

Figure 2 XPS spectra of (**a**) – (**c**) O 1*s*; (**d**) and (**e**) Hf 4*f*; (**f**) and (**g**) Ru 3*p*; of Ru_0.0, Ru_0.5 and Ru_1.0 compounds respectively. The black circles present the experimental data, and the resultant fit from individual deconvoluted peaks (represented in blue, pink, green and cyan colours)) is presented by solid red line.

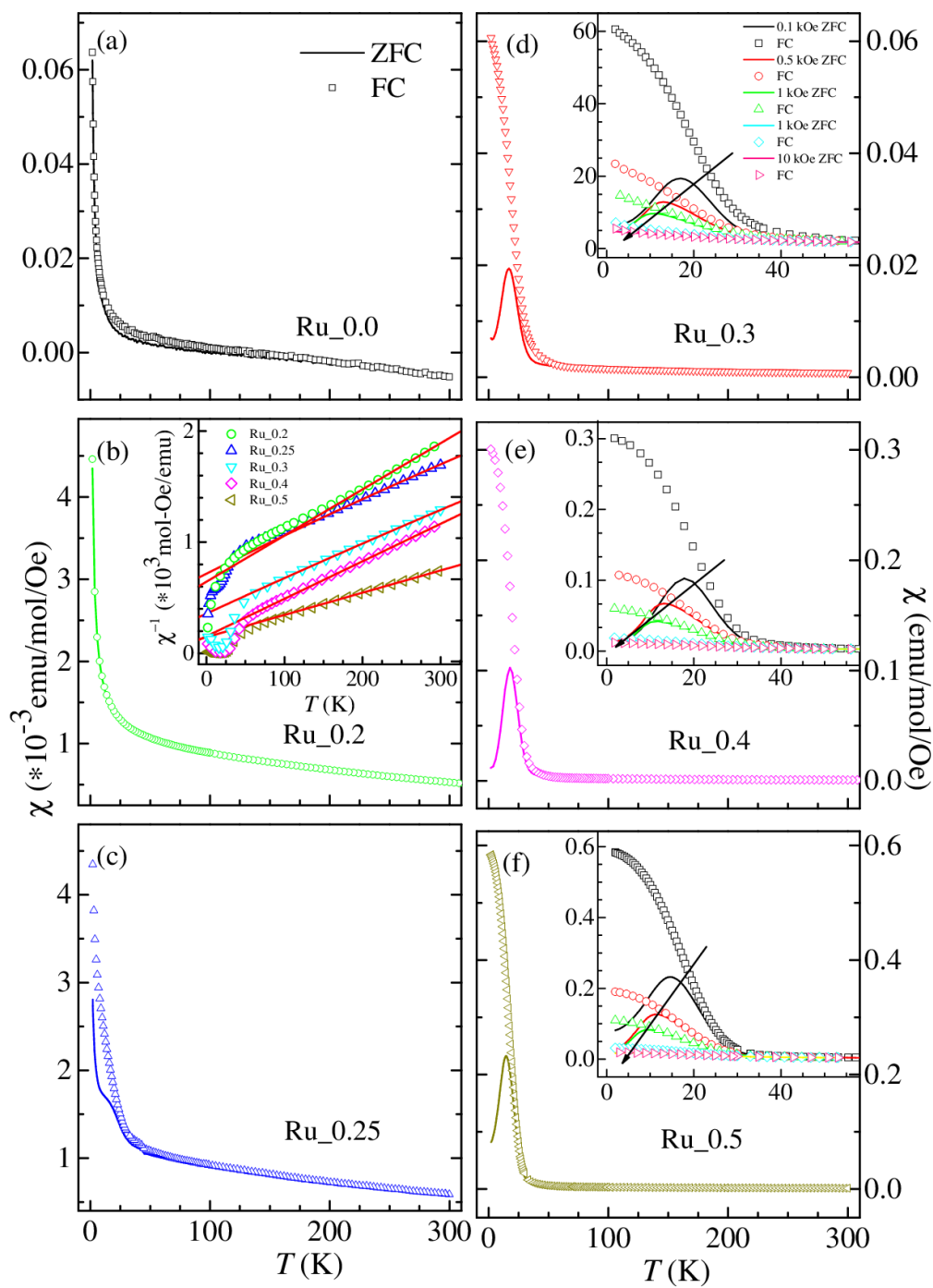

Figure 3 Temperature dependent dc χ under ZFC and FC protocols at 100 Oe for (**a**) Ru_0.0 (**b**) Ru_0.2 (**c**) Ru_0.25 (**d**) Ru_0.3 (**e**) Ru_0.4 (**f**) Ru_0.5 compounds. Inset of (**b**) shows $\chi^{-1}$ vs $T$ with Curie-Weiss fit (solid red line). Insets of (**d**), (**e**) and (**f**) show illustrated view of ZFC – FC curves under different magnetic fields (0.1 – 10.0 kOe).

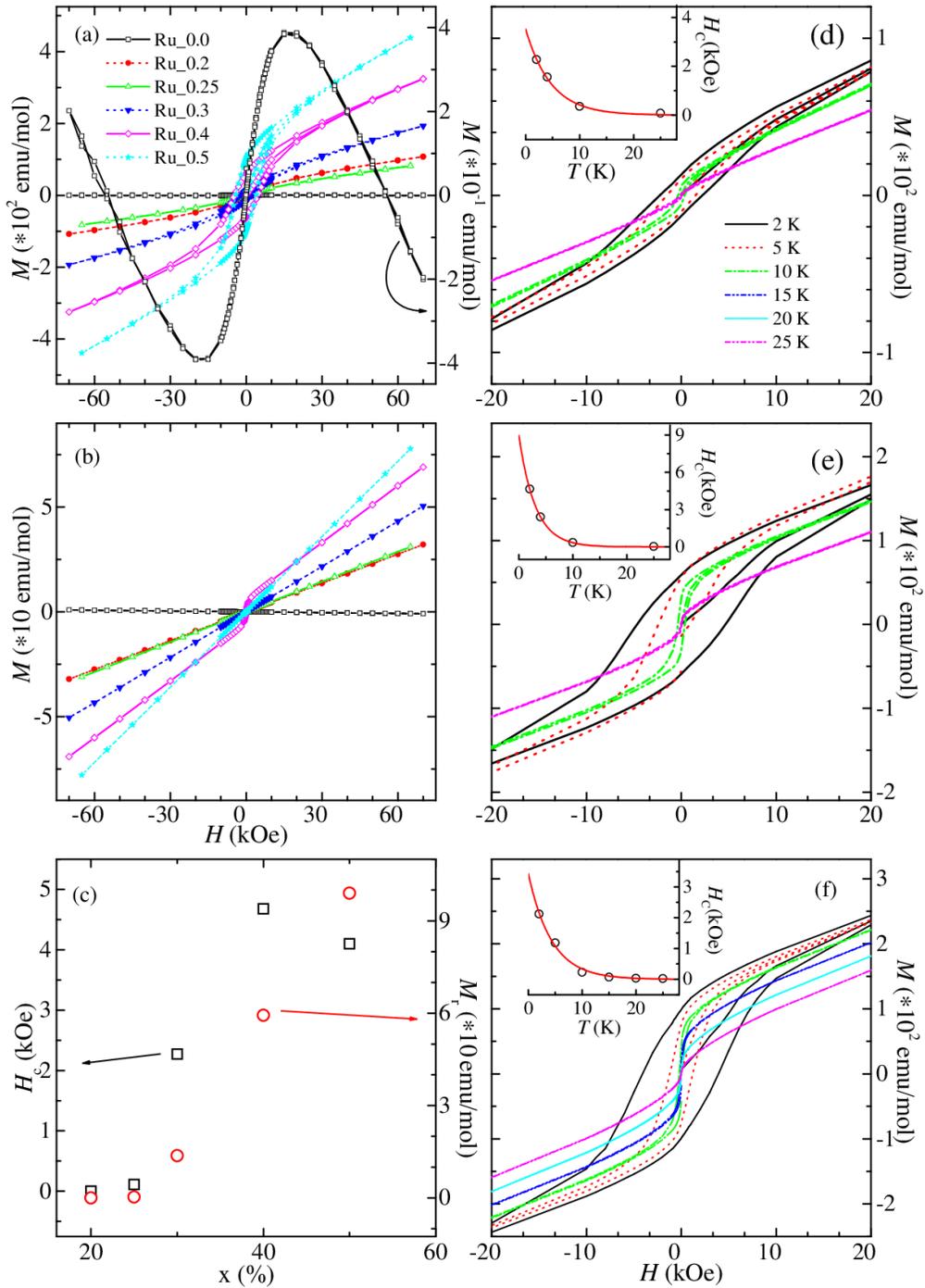

Figure 4 Isothermal *M* (*H*) curves at (**a**) 2 K and (**b**) 300 K for CaHf$_{1-x}$Ru$_x$O$_3$ compounds (**c**) variation of coercive field ($H_c$) and remanent magnetization ($M_r$) at 2 K with increasing Ru concentration. Isothermal *M* (*H*) curves at different *T* are shown for (**d**) Ru_0.3 (**e**) Ru_0.4 and (**f**) Ru_0.5 compounds. Respective insets show the variation of $H_c$ with temperature and solid red line represents the fit using equation (1).

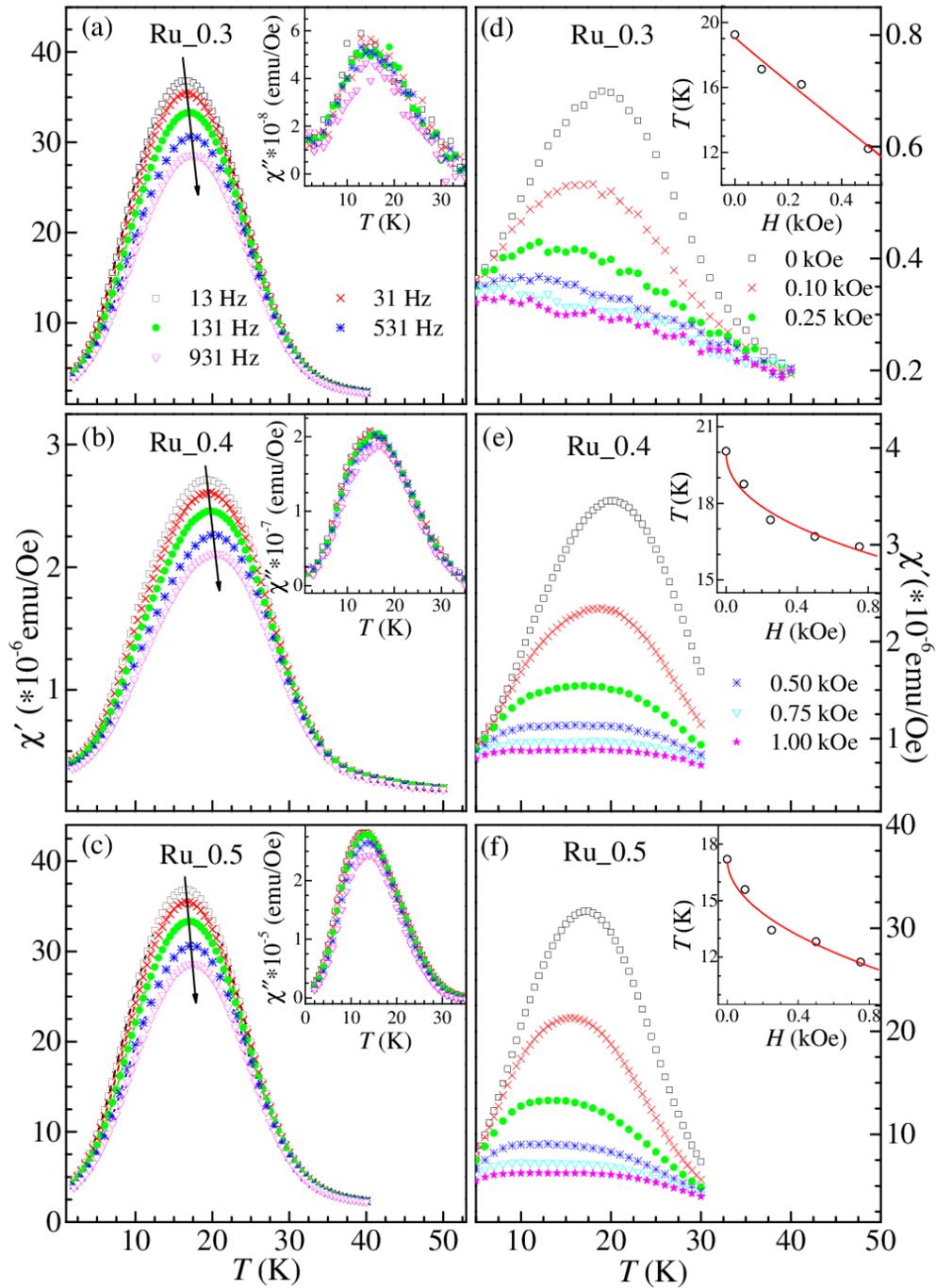

Figure 5 Temperature dependent in–phase ($\chi'$ ($T$)) components of ac susceptibility is shown for (**a**) Ru_0.3 (**b**) Ru_0.4 (**c**) Ru_0.5 and their respective insets show the out–phase components ($\chi''$ ($T$)). Fig (**d**) – (**f**) presents the effect of $H_{dc}$ on $\chi'$ ($T$) and their respective insets present the $H$–$T$ phase diagram with fit according to equation (2).

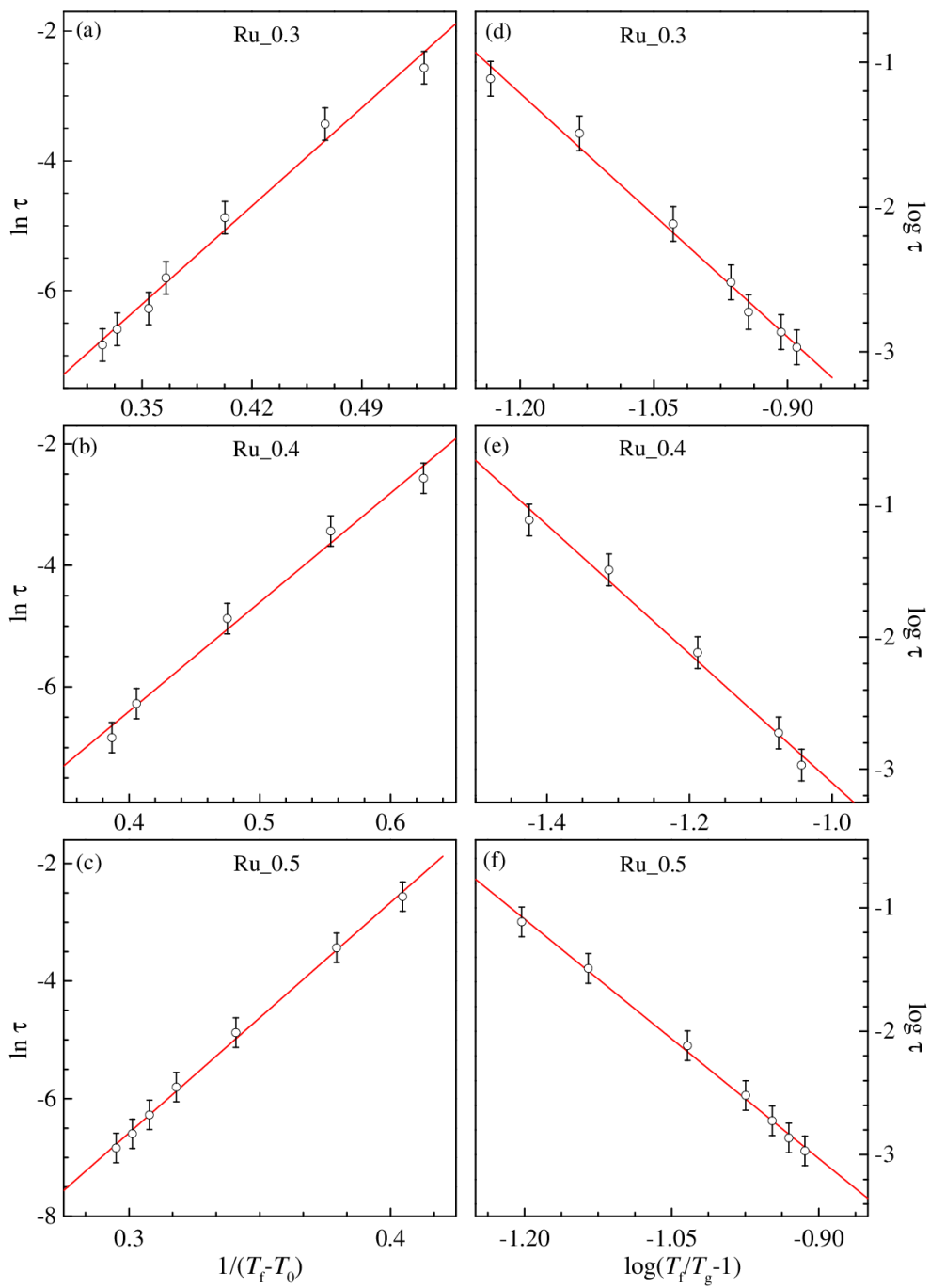

Figure 6 (**a**) – (**c**) Dynamic scaling on frequency dependent $T_f$ by Vogel–Fulcher law fit of relaxation time ($\tau$) as a function of reduced temperature $1/(T_0-T_f)$ using equation (3) and (**d**) – (**f**) Critical slowing down of relaxation time as function of reduced temperature $\log(T_f/T_g-1)$ using Power law equation (4) in (**d**) – (**f**) for Ru_0.30, Ru_0.4, and Ru_0.5 compounds.

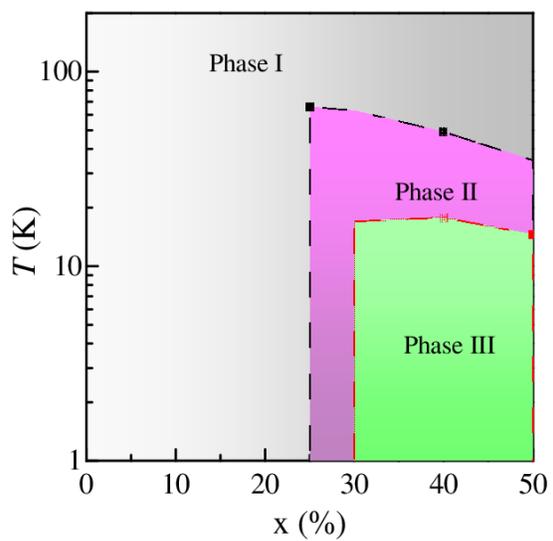

Figure 7 *T*-x phase diagram of CaHf$_{1-x}$Ru$_x$O$_3$ series.

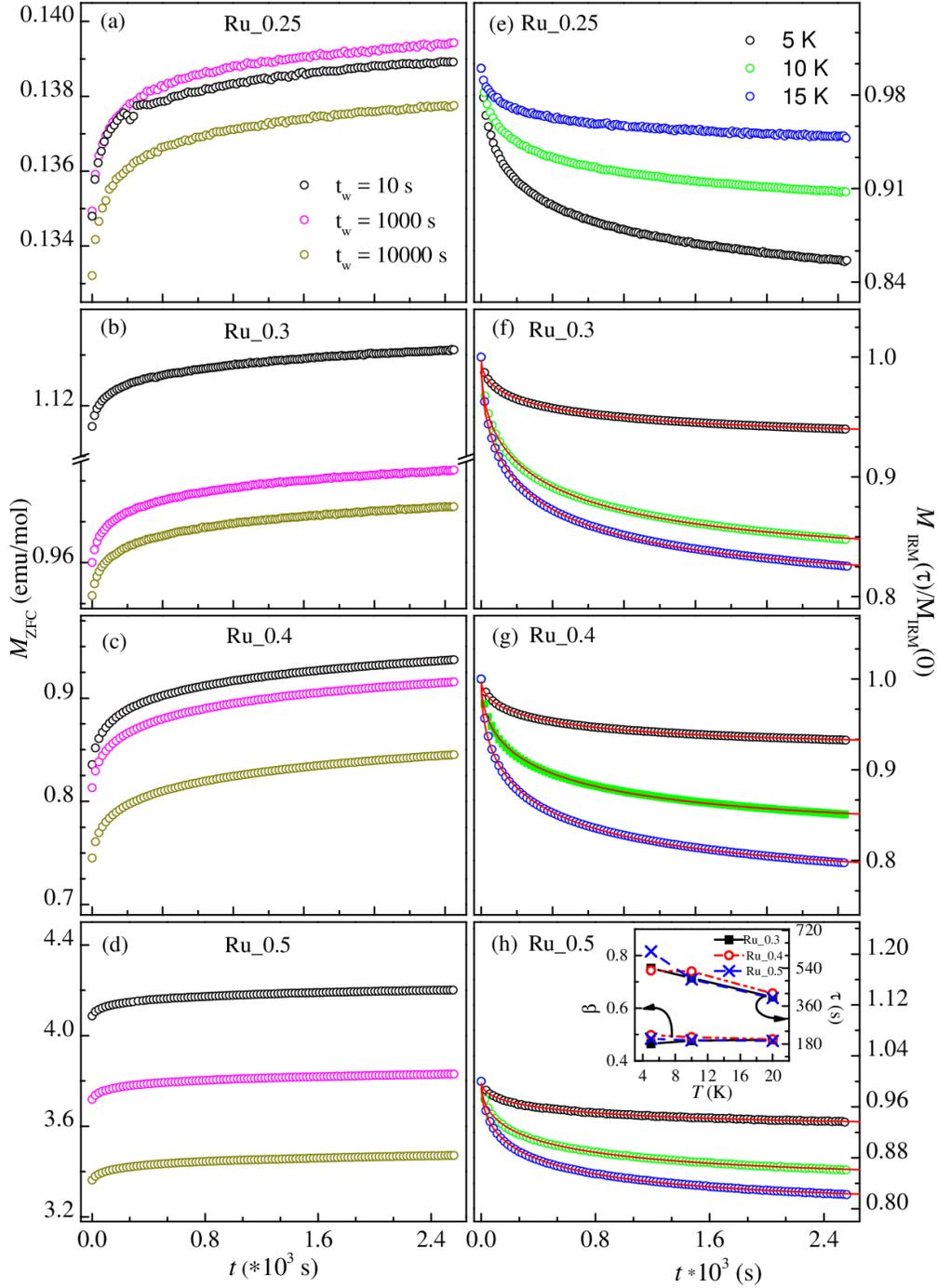

Figure 8 (**a**) – (**d**) shows the aging effect on $M_{ZFC}(t)$ curves at 2 K for different waiting times ($t_w$ = 100 s, 1000 s and 10000 s) (**e**) shows the isothermal remanent magnetization of Ru_0.25 (**f**) – (**g**) shows the isothermal remanent magnetization with fitting (solid red line) using equation (5) at different temperatures for $CaHf_{1-x}Ru_xO_3$ (Ru_0.3 to Ru_0.5) compounds. Inset of (**f**) shows temperature variation of β and τ of these compounds.

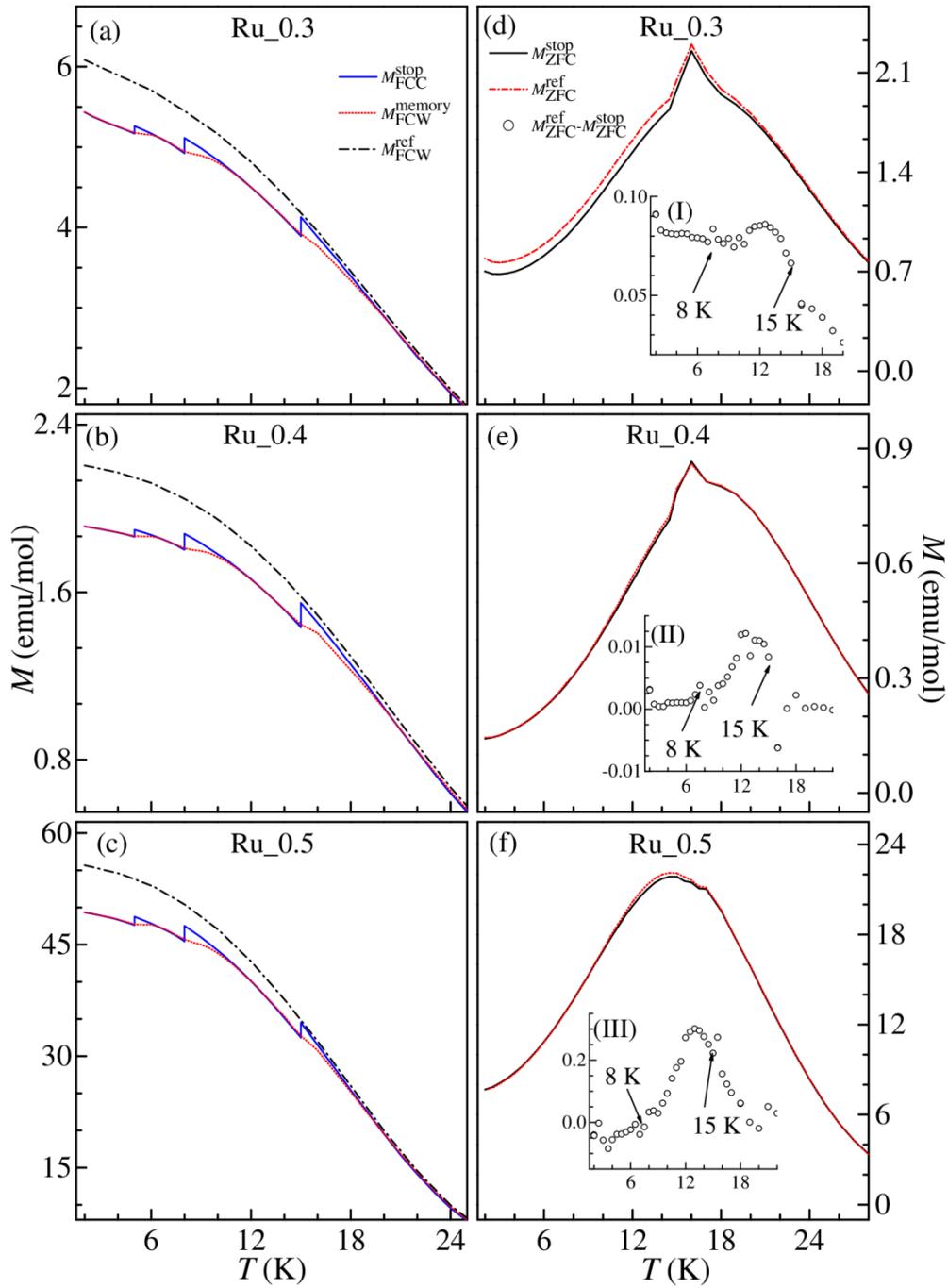

Figure 9 Memory effect as a function of temperature (**a**) – (**c**) under FC protocol (**d**) –(**f**) under ZFC protocol for Ru_0.3, Ru_0.4 and Ru_0.5 compounds. Respective insets of (**d**) – (**f**) show the temperature response of $M^{ref}_{ZFC}$-$M^{stop}_{ZFC}$.

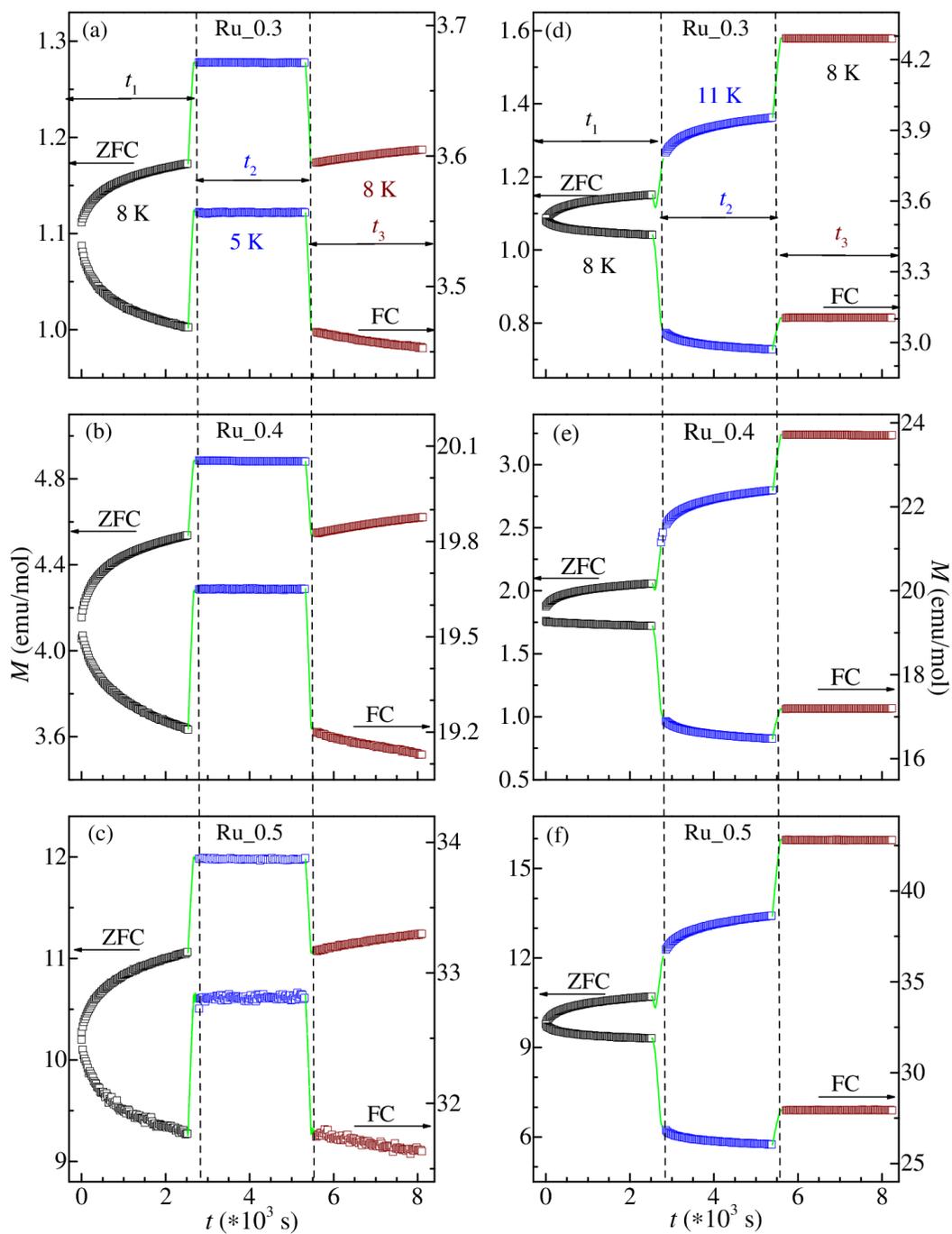

Figure 10 Magnetic relaxations under ZFC and FC protocols during (**a**) – (**c**) cooling $T$ cycle (**d**) – (**f**) heating $T$ cycle for Ru_0.3, Ru_0.4 and Ru_0.5 compounds respectively.

# Supplementary data

# Emergence of low–temperature glassy dynamics in Ru substituted non–magnetic insulator CaHfO$_3$

Gurpreet Kaur and K. Mukherjee

School of Basic Sciences, Indian Institute of Technology Mandi, Mandi 175005, Himachal Pradesh, India

### S1. Raman Spectroscopy

Raman spectroscopy is used to understand the local structure of Ru_0.0, Ru_0.5, and Ru_1.0 compounds. Hence to obtain Raman spectrum, Horiba HR spectrometer is used with 532 nm laser as an excitation source with back scattering geometry. Generally, orthorhombic perovskite compound has 24 Raman active modes ($7A_g + 5B_{1g} + 7B_{2g} + 5B_{3g}$). Due to low intensity or overlapping of modes, we have observed only 10 and 8 Raman active modes in Ru_0.0 and Ru_1.0 compound respectively, from 100–600 cm$^{-1}$ (shown in Fig. S1), which is consistent with literature [1, 2]. The Raman modes observed in Ru_0.5 are a combination of modes obtained in Ru_0.0 and Ru_1.0. This confirms that the crystal structure of the doped compound resembles its end members. The sharpness of peaks reduces when both Ru$^{4+}$ and Hf$^{4+}$ ions are present at the B site. This might be due to the presence of disorder in the doped compound. The nomenclature of Raman modes for Ru_1.0 is obtained using Ref [2]. The peak at 262 cm$^{-1}$ in the Ru_0.0 spectrum corresponds to $A_g(3)$ peak (~ 268 cm$^{-1}$), as observed in Ru_1.0. This peak shifts to 253.5 cm$^{-1}$ in Ru_0.5 compound. This implies that the distortion in BO$_6$ octahedra is increased due to the presence of both cations, Ru$^{4+}$ and Hf$^{4+}$, at B–site of this perovskite.

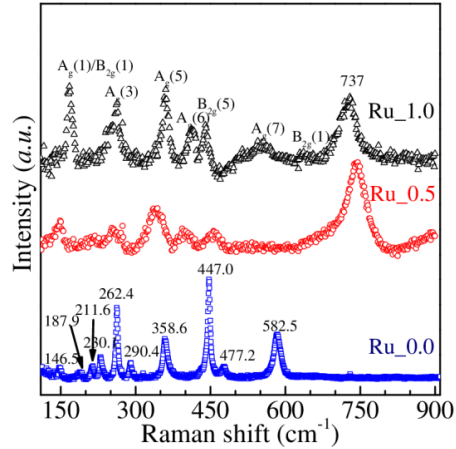

Figure S1 Raman spectrum for Ru_0.0, Ru_0.5 and Ru_1.0 compounds at room temperature.

## S2. DC and AC Magnetisation of $CaRuO_3$

Fig. S2 shows temperature and magnetic field dependent dc magnetisation of $CaRuO_3$ (Ru_1.0). The bifurcation between the ZFC and FC curves (Fig. S2 (a)) in this compound occurs at 85 K, which is consistent with previous studies [3, 4]. The linear behaviour of $M$ ($H$) curves at 300 and 2 K (Fig. S2 (b)), and the negative value of $\theta$ (= -168.47±0.61) obtained from modified Curie–Weiss law (inset of Fig. S2 (a)) indicates to the presence of dominant antiferromagnetic interactions. The value of magnetic moment obtained from Curie–Weiss constant (2.96 $\mu_B$) is close to the moment of $Ru^{4+}$ ion in its low spin ($S = 1$) state. To further explore the low temperature magnetic state ac susceptibility as function of temperature at different frequencies has been performed (shown in Fig. S3 (c)). No frequency or temperature dependent anomaly has been obtained indicating the absence magnetic ordering in this compound. This has been further confirmed by the heat capacity data (Fig. S4).

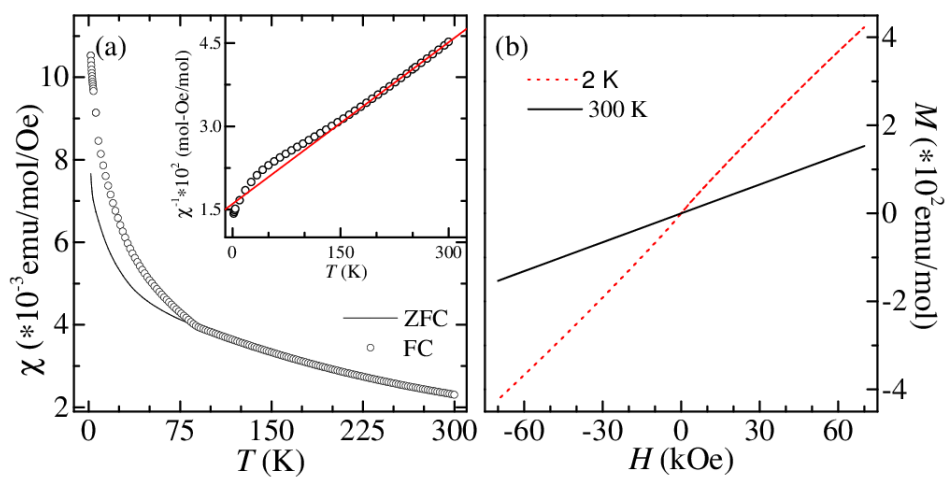

Figure S2 (**a**) Temperature and (**b**) magnetic field dependent dc magnetisation of Ru_1.0 compound. Inset shows the modified Curie–Weiss fit on $\chi^{-1}(T)$.

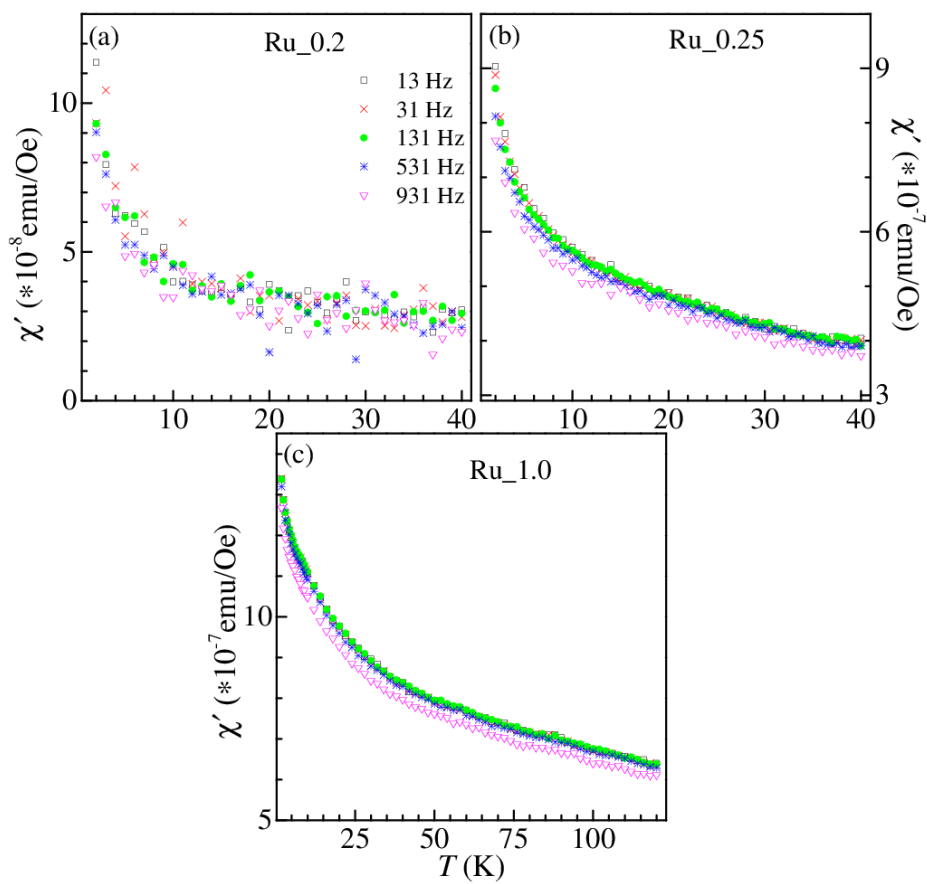

Figure S3 Temperature dependent in–phase component ($\chi'(T)$) of ac susceptibility at different frequencies for (**a**) Ru_0.2, (**b**) Ru_0.25, and (**c**) Ru_1.0 compounds.

## S3. Heat capacity

To further investigate the physical properties in this series, temperature dependent heat capacity ($C(T)$) for selected compounds (i.e. for Ru_0.0, Ru_0.25, Ru_0.4, and Ru_1.0) is measured from 2–200 K in 0 Oe (shown in Fig. S4 (a)). Inset Fig. S4 (a) shows the $C/T$ vs $T$ of these compounds in 2–15 K range. The nonexistence of any anomaly in temperature dependent $C$ and $C/T$ in compounds Ru_0.0 and Ru_0.25 indicate the absence of magnetic ordering in these compounds. The Ru_0.4 compound shows a sharp peak at 7.31 K far below $T_f$ and this anomaly is field independent (not shown here). Also, no kind of long–range magnetic ordering was observed in static as well as dynamic magnetic susceptibility, which abolishes its magnetic origin. Thus, this anomaly might arise from structural phase transition. The Ru_1.0 compound shows an upturn below 7 K in $C/T$ vs $T$ and the value of $C/T$ approaches 81 mJ/mol–K$^2$ as $T$ approaches to zero, consistent with Ref [3]. It has also been noted that the value of heat capacity at low temperatures increases with increasing Ru content in the system. This might happen due to the magnetic contribution of Ru$^{4+}$ cations.

To find the non–magnetic contribution, we have fitted the zero–field heat capacity ($C(T)$) from 100–200 K by the following equation.

$$C(T) = \gamma T + m * 3Nk \left(\frac{T}{\theta_D}\right)^3 \left(3 * \int_0^{\frac{\theta_D}{T}} \frac{x^4 e^x}{(e^x-1)^2} dx\right) + (1-m) * 3NR \left(\frac{\theta_E}{T}\right)^2 \left(\frac{e^{\frac{\theta_E}{T}}}{\left(e^{\frac{\theta_E}{T}}-1\right)^2}\right) \quad (1)$$

where $x = \frac{\hbar \omega_D}{k_B T}$

The first term accounts for the electronic contribution while the second and third terms for lattice contribution to the heat capacity at high temperature. The parameters $\gamma$, $\theta_D$ and $\theta_E$ are the Sommerfeld coefficient, Debye temperature, and Einstein temperature respectively. The values of $m$ and ($1-m$) give information about the contribution of the Debye and Einstein model to the lattice part of heat capacity, respectively. The value of Sommerfeld coefficient ($\gamma = 18$ mJ/mol–K$^2$) and Debye temperature ($\theta_D = 431.68$ K) for Ru_1.0 are close to the values reported in ref [37, 39], with Einstein temperature, $\theta_E = 897.73$ K. The values of $\gamma$, $\theta_D$ and $\theta_E$ for other compounds are tabulated in Table S1. As we move from insulating Ru_0.0 to metallic Ru_1.0, the electronic contribution to heat capacity increases thus $\gamma$ increases from zero to 18 mJ/mol–

$K^2$. Also, the $\theta_D$ and $\theta_E$ decreases while replacing Ru by Hf. To find the magnetic contribution at low temperature, we have subtracted the fitted non–magnetic part by equation (1) from the total heat capacity as shown in Fig. S4 (b). No anomaly around $T_f$ is observed in $C$ vs $T$ but a broad peak at ~ 50 K in $C_m$ vs $T$, have been reported in well–known magnetic glass system. Maximum entropy change calculated for Ru_0.25, Ru_0.4 and Ru_1.0 compositions using $\Delta S_m = \int \frac{C_m}{T} dT$ is much smaller from the expected value Rln(2S+1), unlike the case of long range ordered systems [5]. Such significant reduction in entropy change indicates the presence of multiple degenerate states at low temperature [6,7]. It has also been noted that the magnetic contribution increases with the concentration of magnetic ion $Ru^{4+}$ in the system.

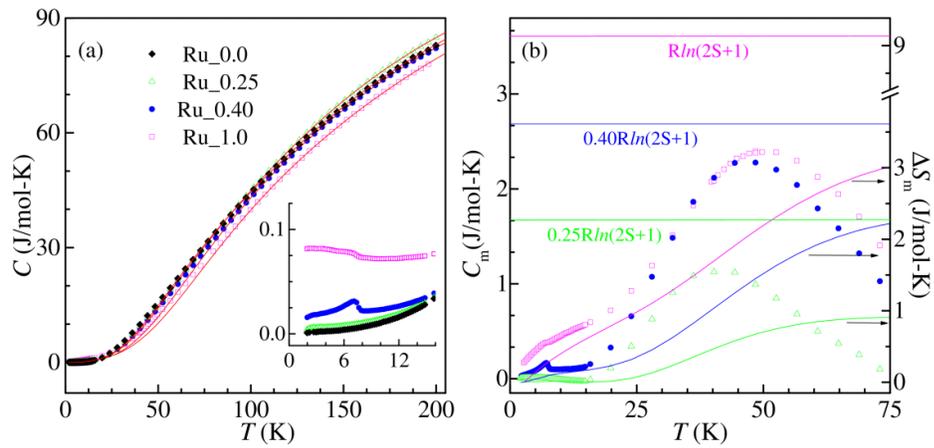

Figure S4 (**a**) Temperature dependent heat capacity at 0 Oe for selected compounds Ru_0.0, Ru_0.25, Ru_0.4 and Ru_1.0, with red solid line fit by equation (5), inset shows zoomed view of low temperature heat capacity. (**b**) shows $C_m$ vs $T$ for Ru_0.25, Ru_0.4 and Ru_1.00 compounds.

Table S1 Parameters obtained from equation (1) fitted on temperature dependent heat capacity at zero field.

| Compounds | γ (mJ/mol–K²) | $\theta_D$ (K) | $\theta_E$ (K) | m |
|---|---|---|---|---|
| Ru_0.0 | 0 | 328.4 (3) | 671.3 (5) | 0.537 |
| Ru_0.25 | 1 | 339.6 (2) | 649.7 (3) | 0.553 |
| Ru_0.4 | 2.6 | 389.5 (2) | 759.0 (0) | 0.640 |
| Ru_1.0 | 18 | 431.7 (3) | 897.0 (0) | 0.650 |